\documentclass[12pt,aps,amssymb,prb,preprint]{revtex4}
\usepackage{graphicx}

\begin{document}

\title{Generalized-Ensemble Simulations\\ of the Human Parathyroid 
       Hormone Fragment PTH(1-34)}   

\author{Ulrich H.E. Hansmann \footnote{
                        ph.: 906-487-2909; fax: 906-487-2933;
                        hansmann@mtu.edu}}
\affiliation{Department of Physics, Michigan Technological University,
         Houghton, MI 49931-1295, USA}
\date{\today}
\begin{abstract}
A generalized-ensemble technique, multicanonical sampling, is 
used to study the folding of a 34-residue human parathyroid
hormone  fragment.  An all-atom model of the peptide is employed
and the protein-solvent interactions are approximated by an
implicit solvent. Our results demonstrate that generalized-ensemble 
simulations are well suited  to sample low-energy structures 
of such large polypeptides. Configurations with a root-mean-square 
deviation (rmsd) to the crystal structure of less than one \AA\ 
are found. Finally, we discuss  limitations of  our
implicit solvent model.
\end{abstract}
\maketitle

\newpage
\section{Introduction}

The successful  deciphering of the human genome has aggravated
an old challenge in protein science: for most of the resolved
 protein sequences we do not know the corresponding structures and 
functions. Computer experiments offer one way to evaluate the
sequence-structure relationship but are extremely difficult for
realistic protein models where interactions among all atoms
are taken into account. The complex form of the intramolecular forces
and of the interaction with the solvent, containing both repulsive and
attractive terms, leads to a very rough energy landscape with a huge number
of local minima. These minima are separated by  energy barriers
that are much higher than the typical thermal energy of a protein 
(of order $k_BT$)  at room temperature. Hence, simple canonical
Monte Carlo or molecular dynamics simulations will get trapped in a
local minimum and often not thermalize within a finite amount of available
CPU time.  While this multiple minima problem does not necessarly  
inhibits molecular dynamics or Monte Carlo studies of peptides and 
proteins \cite{Calfisch1,Kol}, it restricts calculation of accurate
{\it thermodynamic} averages to small peptides \cite{Chow}.

A number of novel simulation techniques  have been proposed for 
overcoming this multiple-minima problem (for a  review, 
see Ref.~\onlinecite{HO98g}).  Important recent examples can be
found in Ref.~\onlinecite{Brooks2002,Elber}. Another example is
parallel tempering, also known as replica exchange method 
and introduced to protein science in Ref.~\onlinecite{H97f},
that has become increasingly popular over the last few 
years \cite{Angel,LHH,Calfisch2}. Parallel tempering is only one
example of a class of new and sophisticated algorithms
commonly summarized as {\it generalized-ensemble} methods \cite{MyReview}.
In this article, we are concerned with another generalized-ensemble
technique, multicanonical  sampling \cite{MU}, 
that was first applied to the protein-folding
problem in Ref.~\onlinecite{HO1}.  Its usefullnes for calculating reliable
thermodynamic quantities at low temperatures has been established
 for small peptides  of up to $\approx 20$ residues  \cite{HO1,HO98c,HA02}. 
However,  stable domains in proteins  consists usually of 40-200 amino
acids. Hence, it is necessary to  evaluate the effectiveness of this
approach for larger molecules than the peptides  investigated so far.
For this purpose, we have performed generalized-ensemble
simulations  of  the peptide fragment PTH(1-34) corresponding 
to residues 1-34 of human parathyroid hormone \cite{Klaus,Crystal,MABFR}. 

The 84-amino acid human parathyroid hormone is involved in the 
regulation of the calcium level in blood and influences bone 
formation \cite{Potts}. The NH$_2$-terminal 34 residues of
the hormone, further on refered to as PTH(1-34), are 
sufficient for the biological activities of this hormone 
suggesting medical and pharmaceutical applications of the 
peptide \cite{Brommage99}. 
The crystal structure of the peptide has been
resolved at 0.9 \AA\  and resembles a slightly bend long
helix (PDB code 1ET1) \cite{Crystal}. NMR studies of the peptide 
in solution under near physiological conditions (PDB code 1ZWA) 
\cite{MABFR} and in 20 \% trifluorethanol solution (PDB code 
1HPY) \cite{MABFR}
rather indicate an ensemble of structures that have in common 
two helices separated by a disordered region. 

In the present article, we try to overcome the problems of
previous simulated annealing simulations of  PTH(1-34) \cite{Okamoto}, 
that did not allow a detailed structure evaluation, by using 
multicanonical  sampling \cite{MU}, one of the most prominent 
generalized-ensemble techniques. 
An all-atom representation of the molecule is employed and 
the intramolecular interactions are 
described by the  ECEPP/3 force field \cite{EC3}. 
The protein-solvent interactions are approximated by the solvent 
accessible surface term of Ooi {\it et al}.\cite{OONS}  Quantities 
such as  the average helicity, number of contacts, average energy, 
and specific heat are calculated.  Our results demonstrate the 
feasibility of generalized-ensemble simulations for  large molecules such
such as PTH(1-34). In addition, they indicate  that with the 
advent of these and other modern search techniques, structure 
prediction of proteins is limited more by current  energy 
functions (and especially solvent approximations) than by the 
simulation algorithms.


\section{Methods}
Our research into the thermodynamics of PTH(1-34)  is based on a detailed,
all-atom representation of that peptide. The interactions between the
atoms are described  by a standard force field, ECEPP/3 \cite{EC3}  
(as implemented in the program package SMMP \cite{SMMP}), and are given by
the sum of the electrostatic term $E_{C}$, the Lennard-Jones energy $E_{LJ}$, 
hydrogen-bond term $E_{HB}$ for all pairs of atoms in the peptide
together with the torsion term $E_{tor}$ for all torsion angles:
\begin{eqnarray}
E_{ECEPP/2} & = & E_{C} + E_{LJ} + E_{HB} + E_{tor},\\
E_{C}  & = & \sum_{(i,j)} \frac{332q_i q_j}{\epsilon r_{ij}},\\
E_{LJ} & = & \sum_{(i,j)} \left( \frac{A_{ij}}{r^{12}_{ij}}
                                - \frac{B_{ij}}{r^6_{ij}} \right),\\
E_{HB}  & = & \sum_{(i,j)} \left( \frac{C_{ij}}{r^{12}_{ij}}
                                - \frac{D_{ij}}{r^{10}_{ij}} \right),\\
E_{tor}& = & \sum_l U_l \left( 1 \pm \cos (n_l \chi_l ) \right).
\label{ECEPP/2}
\end{eqnarray}
Here, $r_{ij}$ (in \AA) is the distance between the atoms $i$ and $j$, 
$\chi_l$ is the torsion angle for the chemical bond $l$ and $n_l$ 
characterizes its symmetry. The charges and
force field parameters $q_i,A_{ij},B_{ij},C_{ij}, D_{ij},U_l$
were calculated from crystal structures of amino acids using  semi-empirical
methods.  The dielectricity constant is set to $\varepsilon=2$, its 
common value in ECEPP calculations. Since the bond lengths are fixed in ECEPP,
the backbone torsion angles $\phi, \psi, \omega$ and the side chain 
torsion angles $\chi$ are the true degrees of freedom. Hence, in a 
Monte Carlo (MC) sweep   single angles are updated sequentially  by the
Metropolis algorithm.   The protein-water interactions  are
approximated by a solvent-accessible surface term 
\begin{equation}
\label{sigma}
E_{solv}=\sum_i\sigma_iA_i~,
\label{Esol}
\end{equation}
where  $A_i$ is the solvent accessible surface area 
of the $i-th$ atom  in the present configuration, and $\sigma_i$  
the  solvation parameter for the atom $i$.  We  choose the 
solvation parameter set OONS of Ref.~\cite{OONS} that is often  
used together with the ECEPP force field. The potential
energy of the solvated molecule is then given by
\begin{equation}
E_{tot} = E_{ECEPP/3} + E_{solv}
\end{equation}
 
In such a detailed protein model, the various competing interactions 
lead model to an energy landscape with a multitude of local minima 
separated by high-energy barriers.  Canonical Monte Carlo or 
molecular dynamics simulations will likely   get trapped in one of these
minima and not thermalize within the available CPU time. Only recently,
with the introduction of new and sophisticated algorithms such as
{\it generalized-ensemble} techniques \cite{MyReview} was it
possible to alleviate this problem in  protein simulations \cite{HO1}.
For simulating PTH(1-34) we have chosen one most commonly used 
generalized-ensemble technique, multicanonical sampling  \cite{MU}.

The multicanonical algorithm \cite{MU} assigns a 
weight $w_{mu} (E)\propto 1/n(E)$ to conformations with energy $E$.
Here, $n(E)$ is the density of states.  A  simulation with this weight
leads to a uniform distribution of energy:
\begin{equation}
  P_{mu}(E) \,  \propto \,  n(E)~w_{mu}(E) = {\rm const}~.
\label{eqmu}
\end{equation}
Thus, the simulation generates a 1D random walk in the energy space,
allowing itself to escape from any  local minimum.
Since a large range of energies is sampled,   re-weighting  \cite{FS} 
allows one to  calculate thermodynamic quantities over a wide range of 
temperatures $T$ by
\begin{equation}
<{\cal{A}}>_T ~=~ \frac{{\int dx~{\cal{A}}(x)~w^{-1}(E(x))~
                 e^{-\beta E(x)}}}
              {{\int dx~w^{-1}(E(x))~e^{-\beta E(x)}}}~,
\label{eqrw}
\end{equation}
where $x$ stands for configurations, $E(x)$ for its total potential
energy $E(x) = E_{ECEPP/3} (x) + E_{solv}(x)$ and $\beta$ for the inverse 
temperature, $\beta = 1/k_B T$.

Unlike in constant temperature simulations the weights are 
not {\it a priori} known in multicanonical simulations. In fact,
knowledge of the exact weights is equivalent to obtaining the 
density of states $n(E)$, i.e., solving the system. However,
for a numerical simulations estimators are sufficient as long as these
do not deviate not too much from $n^{-1}(E)$. This is because the
same weights that are used for the simulation appear also in the
re-weighting procedure of Eq.~\ref{eqrw}.  In the present study, we 
calculate these estimators from a preliminary simulated annealing 
run  of 80,000 MC-sweeps using the method described in Ref.~\onlinecite{H97}. 
All thermodynamic quantities are then estimated from one production
run of $1,000,000$ sweeps, starting from a random initial configuration
and after discarding  $10,000$ sweeps  for thermalization. 
We store in every fifth sweep for further analysis various 
physical quantities and the dihedral angles of the current 
configuration. Our error bars are estimated by dividing 
this time series of data into 8 bins of each $125,000$ sweeps.

\section{Results and Discussion}
We start our analysis by calculating    thermodynamic averages of 
the  intramolecular energy $<E_{ECEPP/3}>(T)$ and the solvation energy
$<E_{SOLV}>(T)$. Both quantities and the resulting total energy
$<E_{TOT}>(T) = <E_{ECEPP/3} + E_{SOLV}>(T)$ are displayed as function
of temperature in Fig.~1. The thermal behavior of the peptide is
characterized by a competition between intramolecular and solvation
energy. While $<E_{ECEPP/3}>(T)$ decreases with decreasing temperature,
$<E_{SOLV}>(T)$ increases toward lower temperatures. The interplay of
both terms leads to two temperature regimes
that are separated by a steep decrease in the total energy $<E_{TOT}>(T)$.
The corresponding pronounced peak in the specific heat per residue
\begin{equation}
 C(T) = \beta^2 \left(<E^2_{TOT}>(T) - <E_{TOT}>^2(T)\right)/34,
\end{equation}
displayed in the inlet, marks the transition temperature  at
 $T_c = 560\pm 10$ K. 

The structural changes associated with
this transition can be deduced from Fig.~2 where we display
the average helicity $<n_H>(T)$ 
as a function of temperature. Here, we have defined $n_H$ as the number 
of residues whose pair of backbone dihedral angles $(\phi,\psi)$ takes 
values in the range: $(-70^{\circ}\pm 30^{\circ},-37^{\circ}\pm 30^{\circ})$. 
We see from the plot of both quantities that the high-temperature 
(high-energy) region is characterized by configurations with vanishing 
helicity ($\approx 10$\%) while at low temperatures (and, correspondingly,
 low energies)
configurations dominate that are almost completely helical 
($\approx 90$\% of the residues are part of a helix). The pronounced
peak at $T_c$ in the susceptibility (per residue)
\begin{equation}
\chi(T) = \left( <n^2_H> - <n_H>^2 \right)/34,
\end{equation} 
shown in the inlet, is further proof for the sharp transition between
low-energy helical states and high-energy disordered coil states. 
Associated with this helix-coil transition is also a  decrease
in the solvent accessible volume $<V>(T)$ as  calculated by
the double cubic lattice method \cite{ELASS} (data not shown). 
Above $T_c$, the average volume is $<V>\approx 10000$ \AA$^3$, while
below $T_c$ the volume is reduced to $<V>\approx 9000$ \AA$^3$. 

The modest decrease in $<V>$ together with the large value of the helicity
$<n_H>(T)$ indicate that below $T_c$ a single elongated helix  is the
dominant structure for PTH(1-34). Indeed, we find in our simulation
as lowest-energy state an elongated helix  with 31 residues part
of the helix. This structure, shown in Fig.~3b, has not only the 
lowest {\it total} energy ($E_{TOT} = -277.8$ kcal/mol), but also
the lowest {\it intramolecular} energy: $E_{ECEPP/3} = -136.5$ kcal/mol. 
It is very similar to the crystal structure of PTH(1-34) (PDB code 1ET1, 
displayed in Fig.~3a) where also 31 residues are part of an $\alpha$-helix
and whose energy after regularization with the program FANTOM \cite{FANTOM}
is $E_{TOT} = -277.9$ kcal/mol ($E_{ECEPP/3} = -187.0$ kcal/mol). While
our numerically determined structure  has a slightly larger 
solvent-accessible surface area ($A= 3860$ \AA$^2$) than the crystal
structure ($A=3410$ \AA$^2$), it has 95\% of all native contacts formed,
i.e. 95 \% of the contacts between residues found in the crystal structure
exist also in our lowest-energy configuration. Here, we consider
two residues in contact if the distance between their C$_{\alpha}$-atoms
is less than 8 \AA, and the two residues are neither neighbor nor 
next-nearest neighbor in the peptide chain. Given that almost all
native contacts are formed in the lowest-energy structure, it is not 
surprising that  the 
root-mean-square deviation (rmsd) between the two structures is
only $0.8$ \AA\ for backbone atoms ($2.3$ \AA\ when all heavy atoms
are taken into account). For comparison, the crystal structure
of PTH(1-34) (1ET1) itself was solved at $0.9$ \AA\ resolution \cite{Crystal}.
We remark that recent structure determinations of the similar sized
villin headpiece subdomain HP-36, a 36-residue peptide,
by Energy Landscape Paving \cite{HW} and parallel tempering \cite{LHH}
were restricted to an accuracy of $\approx 6$ \AA. We believe that
the higher accuracy of our PTH(1-34) results does not indicate any 
advantage of multicanonical sampling over the above methods but is 
rather due to the simpler geometry of PTH(1-34).

At $T=300$ K, our lowest energy structure appears with a frequency
of $(99\pm 0.5)$\%, i.e.  almost all observed configurations
resemble the crystal structure. The predominance of this structure
is also reflected in the well-developed funnel  in Fig.~4
where we display the projection of the free-energy 
landscape at $T=300$ K on the number of native  contacts. The
free energy decreases rapidly with increasing number of native
contacts.  No indications for competing local minima that
could act as traps are observed indicating  a rather smooth funnel.
On the other hand, at the transition temperature $T_c = 560$ K, 
the free energy landscape (displayed in the inlet) is flat and
configurations  with small  number of native contacts coexist with
such that have many native contacts.

However, while the crystal structure is in our simulation the 
dominant configuration at $T=300$ K,  it differs from the set 
of NMR-structures found at room temperature. In near-physiological
solution,  one observes instead two helices separated by a 
disordered and flexible  region. We show in Fig.~3c  as an 
example one of the resolved solution configurations (from 
1HPY) \cite{MABFR}. The N-terminal helix ranges from
Glu$_4$ to His$_9$ and the C-terminal helix from Ser$_{17}$ to
Gln$_{29}$. Addition of trifluorethanol  reduces hydrophobic
interactions and increases the length of these helices \cite{MABFR}.
Hence, our simulation of PTH(1-34) does
not reproduce the experimental results for that peptide in
solution albeit protein-solvent interactions are considered
in our energy function by an approximate term. Instead, our
simulation favors the crystal structure of the peptide that is
observed in membrane and hydrophobic environments. 

In order to understand in greater detail the relation between 
our simulation results and the NMR experiments, 
we  plot in Fig.~5a for each residue   the free energy
difference $\Delta G_i$ at $T=300$ k between configurations 
with residue $i$  part of an $\alpha$-helix and such where 
that residue is not part of an $\alpha$-helix. The free-energy 
differences are largest for residues Asn$_{16}$ -Lys$_{27}$, 
 and it is for these residues that first
helix formation is observed. A second cluster of residues that have
large free-energy differences are observed between Ile$_5$ and Asn$_{10}$. 
Both regions are separated by residues Leu$_{11}$ - Leu$_{15}$ that
have smaller free energy differences. 
The observed free-energy differences are strongly correlated with differences 
in the (potential) energy $\Delta E_{TOT}$ that together with its two
components $E_{ECEPP/3}$ and $E_{SOLV}$ are plotted in Fig.~5b. Note,
that the variations in the energy differences result from the
$E_{ECEPP/3}$ part, i.e. from the intramolecular interactions. The
corresponding solvent energies favor in general residues that are
not in a helical state but  vary little with the residues. In addition,
their magnitude is so small that it is difficult to distinguish
in the figure between $\Delta E_{TOT}$ and $\Delta E_{ECEPP/3}$.

The position of the two helices in the 
solvent structure of Fig.~3c corresponds to the regions where  
in our simulation the measured free-energy differences and 
potential energy differences  are  large.  In the NMR structures, 
the C-terminal helix is more stable than the N-terminal helix. 
Similarly, we find  larger absolute values of 
$\Delta G_i \approx -14$ kcal/mol 
($\Delta E_{TOT} \approx - 39$ kcal/mol) for residues Asn$_{16}$ 
to Lys$_{27}$ (with the maximal values at Arg$_{20}$: 
$\Delta G = -19.4$ kcal/mol and $\Delta E_{TOT} = -55.1$ kcal/mol)
compared with $\Delta G_i \approx -12$ kcal/mol 
($\Delta E_{TOT} \approx -36$ kcal/mol) for residues Ile$_{5}$ 
to Asn$_{10}$. On the other hand, the flexible region
connecting the two helices  in the NMR structure corresponds to
a region of residues  that have with $\Delta G_i \approx -8$ kcal/mol 
and $\Delta E_{TOT} \approx -25$ kcal/mol  considerably smaller  
free (potential) energy differences. The free-energy differences 
are smallest for Leu$_{11}$ and Gly$_{12}$: $\Delta G_i \approx -6$ kcal/mol 
and and $\Delta E_{TOT} \approx 17$ kcal/mol.  The later result 
is not surprising giving the inherent flexibility of glycine 
(which, however, is part of the helix in our lowest-energy configuration).

The observed variations in the free and potential energy differences
suggest that  we may find at higher temperatures  configurations 
similar to the  NMR structures. This is because the helix will be
de-stabilized with increasing temperature, and more easily
for residues Leu$_{11}$ to Leu$_{15}$ than in the  regions that 
corresponds to the two terminal helices. 
In order to test this conjecture, we show in Fig.~6 two quantities
as function of temperature.  One is the frequency of configurations
that have  a continuous helix stretching at least between Ile$_{5}$ 
and Lys$_{27}$  and are therefore similar to the crystal structure 
of PTH(1-34).  The second quantity is the frequency of configurations 
that have helices stretching at least between Ile$_5$ and His$_{9}$ and
between Ser$_{17}$ and Gln$_{29}$, but are separated in-between
by a non-helical flexible region. Hence, the later quantity measures
the frequency of configurations that are similar to the NMR structure.
A typical example of this group of configurations is displayed in
Fig.~3d. While this conformation is also a local minimum, its 
total energy $E_{TOT} = -201.4$ kcal/mol and 
intramolecular energy $E_{ECEPP/3} = -48.6$ kcal/mol are much
higher than the corresponding  values for the lowest-energy 
configuration of Fig.~3b: $E_{TOT} = -277.8$ kcal/mol and 
 $E_{ECEPP/3} = -135.5$ kcal/mol. On the other hand, its solvation
energy $E_{SOLV} = -153.0$ kcal/mol is lower than that of
Fig.~3b ($E_{SOLV} = -141.3$ kcal/mol) but the differences are
smaller than the one in the ECEPP/3 term.  

Both kind of configurations  appear at the helix-coil
transition temperature $T_c =560(10)$ K. Due to their higher entropy,
configurations that resemble the NMR structures are slightly more 
common for temperatures in the range $520$ K $\le T \le  560$ K
than the ones that are similar 
to the crystal structure leading to a positive free energy 
difference $\Delta G$ that is displayed in the inlet of Fig.~6.  
At $T=520$ K, 43(4)\% of all configurations 
are similar to the NMR structures and 40(5)\% resemble more the
crystal structure. Below $T=520$ K, the intramolecular energy
that favors an extended single helix wins over the higher entropy
of states with two separate helices. The resulting negative
free energy difference $\Delta G$ leads to a decrease in the frequency
of NMR-like structures, and  their contribution is 
less than 1\% at room temperature.

We conjecture  that the above relations hold also in nature.
The elongated helix of the crystal structure is favored by the 
intramolecular energies and is the ground state in potential
energy. Thermal fluctuations lead  to the more flexible configurations
with two helices that  are observed for the soluted peptide. 
In a (hydrophobic) membrane environment or when binding to a receptor 
reduces the entropy of the molecule, PTH(1-34) stays in the 1-helix 
state that also seems to lead to increase its biological 
activity \cite{Crystal}.
 
However,  the energy function in our simulation  overstabilizes
the extended $\alpha$-helix of the PTH(1-34) ground state (the crystal 
structure). Hence, structures that are found in solution with NMR 
experiments appear in our simulation with significant frequency only  
at temperatures more than $200$ K above room temperature. Similarly,
we find with $T_c = 560(10)$ K a helix-coil transition temperature that 
is much higher than  physiological relevant temperature range. These results
clearly point out the limitations of our model. Since our data are 
consistent with what one would expect in a hydrophobic environment
one can conjecture that these limitations of the energy function are 
mainly due to our solvent approximation. The small variation 
in the magnitude of the solvent energy in Figs.~1 and 5 (when compared with
the ECEPP/3 term) suggests that  the OONS term underestimates
the protein-solvent interaction. This maybe due to the fact that 
our implicit solvent model  does not account for the hydrogen bonds 
between the polypeptide and water molecules that in water compete with 
the characteristic intramolecular hydrogen bonding  in an $\alpha$-helix.
As a consequence, $\alpha$-helices are overstabilized, and our energy
function  rather models the peptide in a hydrophobic environment than 
in physiological solution. 
Another problem is that the solvation energy term of Eq.~\ref{Esol}
describes actually a free energy and therefore should change with temperature.
This effect is neglected in the OONS approximation. Taking such a temperature
dependence into account and use of more sophisticated implicit solvent
models would likely improve our results. 
However, we can also not exclude the possibility that the deviation from the
NMR results is not duee to the solvent term bit
that our force field, the ECEPP/3 term that describes the intramolecular
interactions, biases toward helical conformations.

It follows that other potential energy functions and implicit solvent models 
have to be chosen for a simulation of PTH(1-34) in solution. However, the
failure of our energy function to model correctly the solvated molecule 
demonstrates also the advantages  of our approach. Unlike  the earlier simulated
annealing  simulations of Ref.~\onlinecite{Okamoto}   multicanonical sampling
allows one to calculate accurate estimates of the frequency of states and 
other physical quantities at room temperature. In that way, we have been
able to unveil in the present paper the limitations of our energy function. 
At the same time, our results allow us to understand the behavior of our
peptide in a hydrophobic environment (since this is what our energy function 
models) and to reproduce the crystal structure with high accuracy. Hence,
we have shown that multicanonical simulations are not only well-suited for 
simulations of small peptides but also for numerically more
challenging molecules such as the 34-residue peptide PTH(1-34).

\section{Conclusion}

In summary, we have performed all-atom simulations of PTH(1-34), the 
biologically active peptide-fragment 1-34 of the human parathyroid 
hormone. Protein-water interactions are approximated by a
solvent-accessible surface term using the OONS parameter set\cite{OONS}.
Our results rely on multicanonical sampling and demonstrate 
that this technique  allows one  to overcome the multiple minima 
problem in  simulations of this molecule. Unlike earlier numerical
studies \cite{Okamoto} that relied on simulated annealing we find
a  lowest-energy configurations that has a rmsd of only 0.8 \AA\ to 
the crystal structure. While previous applications of 
multicanonical sampling were restricted to small
peptides with less than 20 residues, our results (albeit for
a molecule with a rather simple structure) establish that the 
generalized-ensemble approach is also a useful tool for investigation 
of much larger polypeptides with   $30-40$ residues. 
Applications to even larger molecules seem to be restricted less by the
sampling technique but by the accuracy of the energy function.

\noindent
{\bf Acknowledgments}: \\
Support by a research grant (CHE-9981874) of the National Science 
Foundation is gratefully acknowledged. Part of the described research
was done while I was visiting  the
John von Neumann Institute for Computing (NIC), located in the
FZ J{\"u}lich. I like to thank Prof. Grassberger and the institute for
kind hospitality and for bearing with my  working style.


\vfil

\clearpage
\newpage
{\huge Figure Captions:} 
\begin{description}
\item[Fig.~1] Average intramolecular energy $<E_{ECEPP/3}>$, solvent
              energy $<E_{SOLV}>$ and total energy $<E_{TOT}> = 
              <E_{ECEPP/3} + E_{SOLV}>$ of PTH(1-34) as a function of 
              temperature $T$.  The inlet displays the specific heat $C(T)$
              as a function of temperature $T$. 
              The data are calculated from a multicanonical simulation
              of 1,000,000 sweeps using a solvent accessible surface
              term to approximate protein-water interactions.
\item[Fig.~2] Average helicity $<n_H>$ as a function of temperature $T$.
              The inlet displays the susceptibility $\chi(T)$ as
              a function of temperature $T$. All
              data points are calculated from a multicanonical simulation
              of 1,000,000 sweeps using a solvent accessible surface
              term to approximate protein-water interactions.
\item[Fig.~3] (a) The crystal structure of PTH(1-34) (PDB code 1ET1);
              (b) The lowest-energy conformation of PTH(1-34) as determined
                 from a multicanonical simulation with a solvent
                 accessible surface term;
              (c) Solution structure of PTH(1-34) as determined by NMR 
                 (PDB-code 1HPY).
              (d) One of the configurations resembling the solution
                 structure. Such configurations appear in our simulation 
                 with significant frequency only for
                 temperatures closely to but below the 
                 helix-coil transition temperature $T_c =560$ K.
\item[Fig.~4] Projection of the free-energy landscape ($\Delta G$) of PTH(1-34)
              at $T=300$ K on the number of native contacts $n_{NC}$. The
              same quantity  is plotted for the $T=560$ K (the critical
              temperature) in the inlet. 
\item[Fig.~5] (a) Free energy difference $\Delta G_i$ between
                 configurations with residue $i$ part of an $\alpha$-helix
                 and such where residue $i$ is not in a helical state.
              (b) The corresponding total (potential) energy differences
                 $\Delta E_{TOT}$ ($\square$), intramolecular energy differences
                 $\Delta E_{ECEPP/3}$ ($\circ$) and solvent energy differences
                 $\Delta E_{SOLV}$($x$).
\item[Fig.~6] Frequency of PTH(1-34) configurations ($1H$) that  
              resemble the crystal structure and frequency of 
              states ($2H$) that are similar to the solution structures 
              of the peptide. The free-energy difference $\Delta G$ 
              between both sets of configurations is displayed in the 
              inlet. All quantities are shown as function of temperature
              $T$. All our data points 
              are   calculated from a multicanonical simulation
              of 1,000,000 sweeps using a solvent accessible surface
              term to approximate protein-water interactions.  
\end{description}

\newpage
\thispagestyle{empty}
\cleardoublepage
\begin{figure}
\includegraphics[angle=0,width=0.95\textwidth]{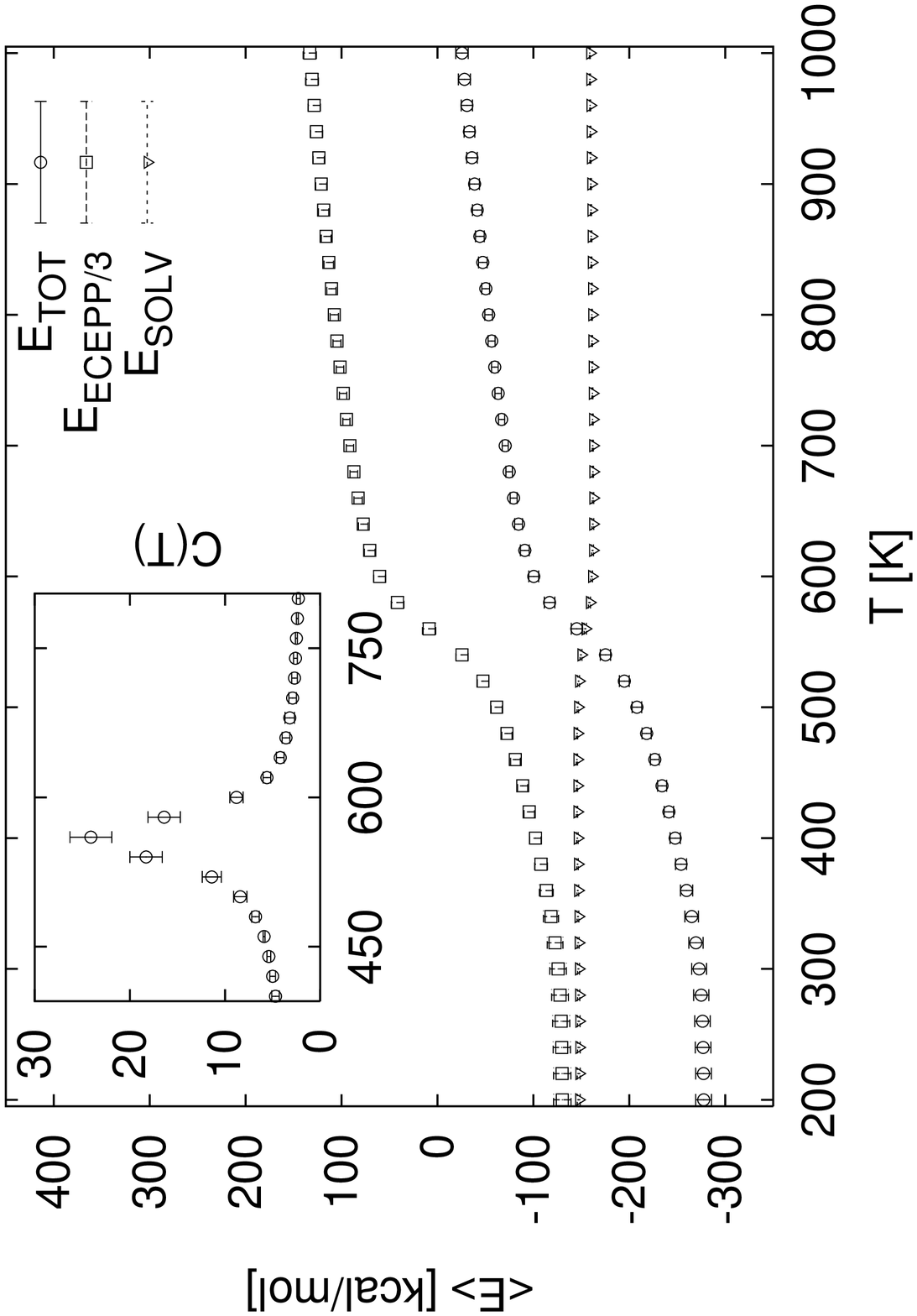}
\label{fig1}
\caption{}
\end{figure}

\newpage
\thispagestyle{empty}
\cleardoublepage
\begin{figure}
\includegraphics[angle=0,width=0.95\textwidth]{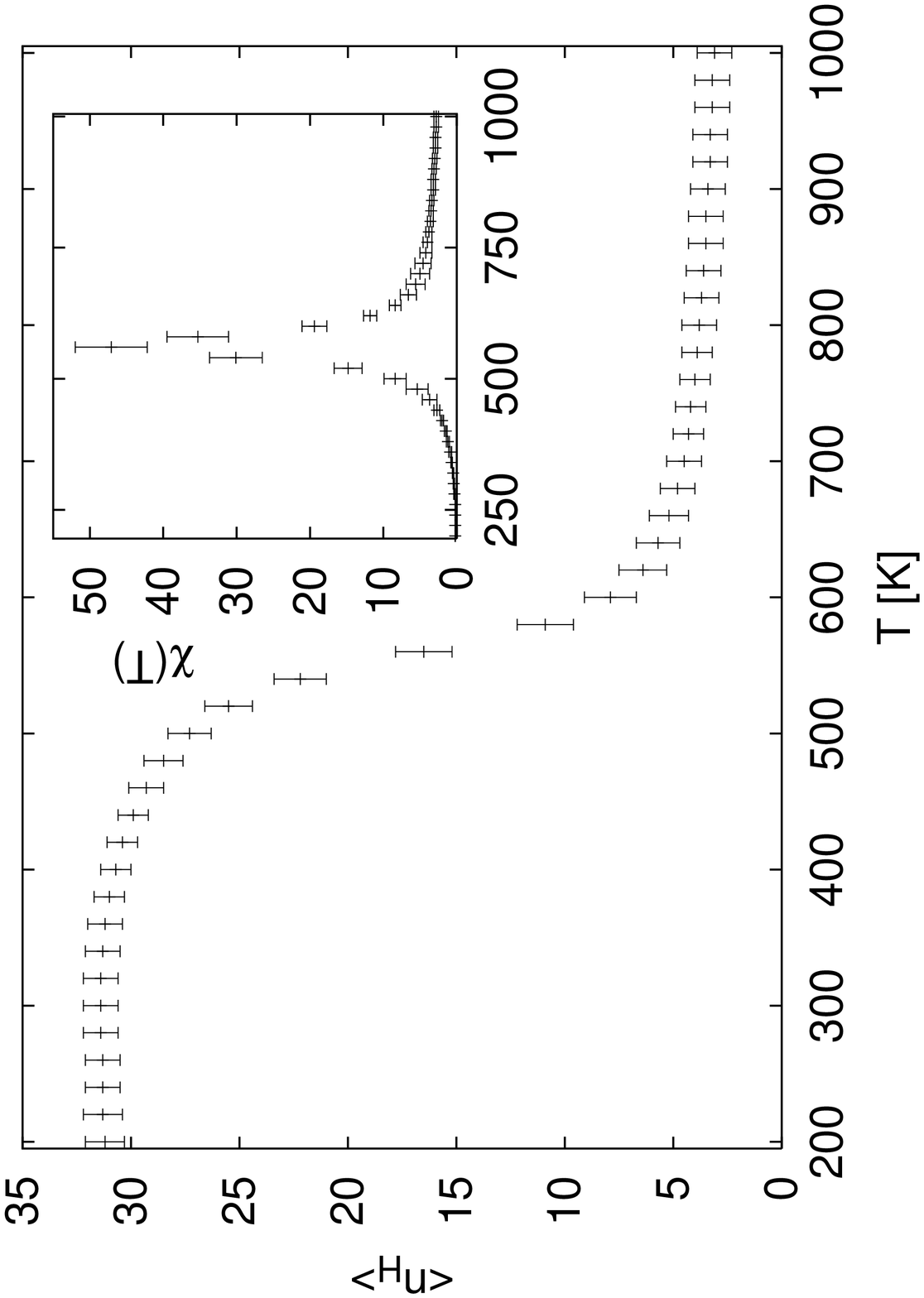}
\caption{}
\end{figure}

\newpage
\thispagestyle{empty}
\cleardoublepage
\begin{figure}
\includegraphics[angle=0,width=0.95\textwidth]{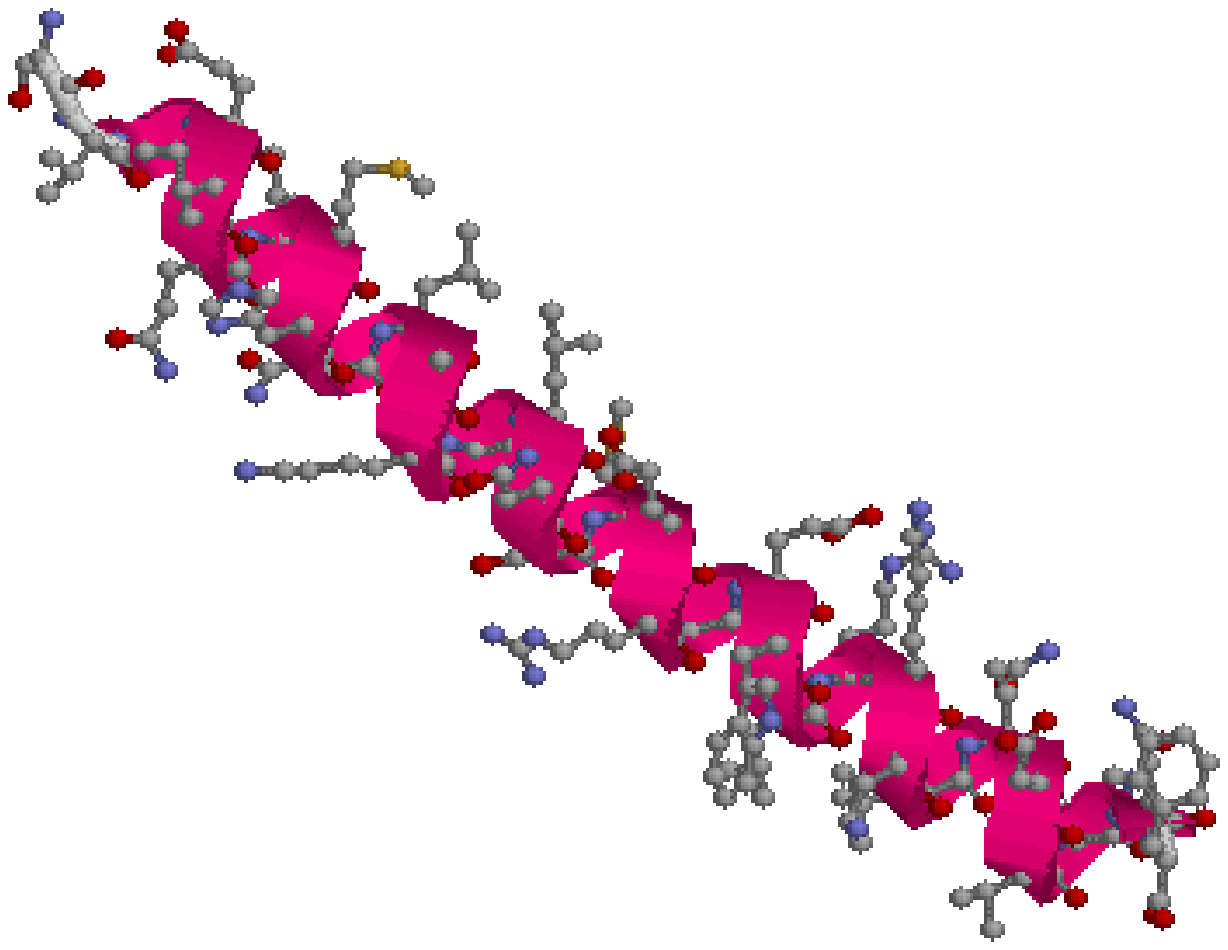}
\caption{}
\end{figure}
\newpage
\thispagestyle{empty}
\cleardoublepage
\begin{figure}
\includegraphics[angle=0,width=0.95\textwidth]{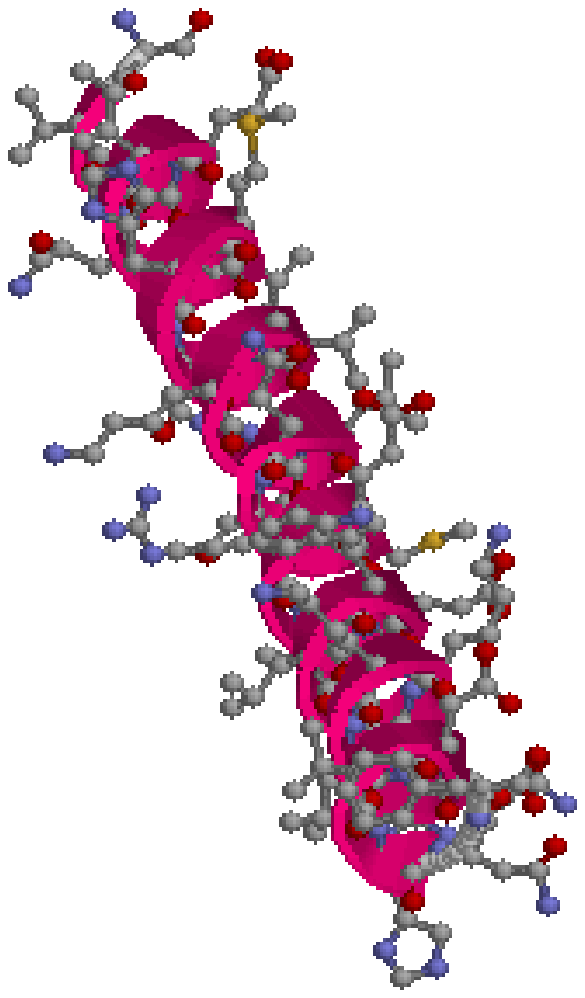}
\caption{}
\end{figure}
\newpage
\thispagestyle{empty}
\cleardoublepage
\begin{figure}
\includegraphics[angle=0,width=0.95\textwidth]{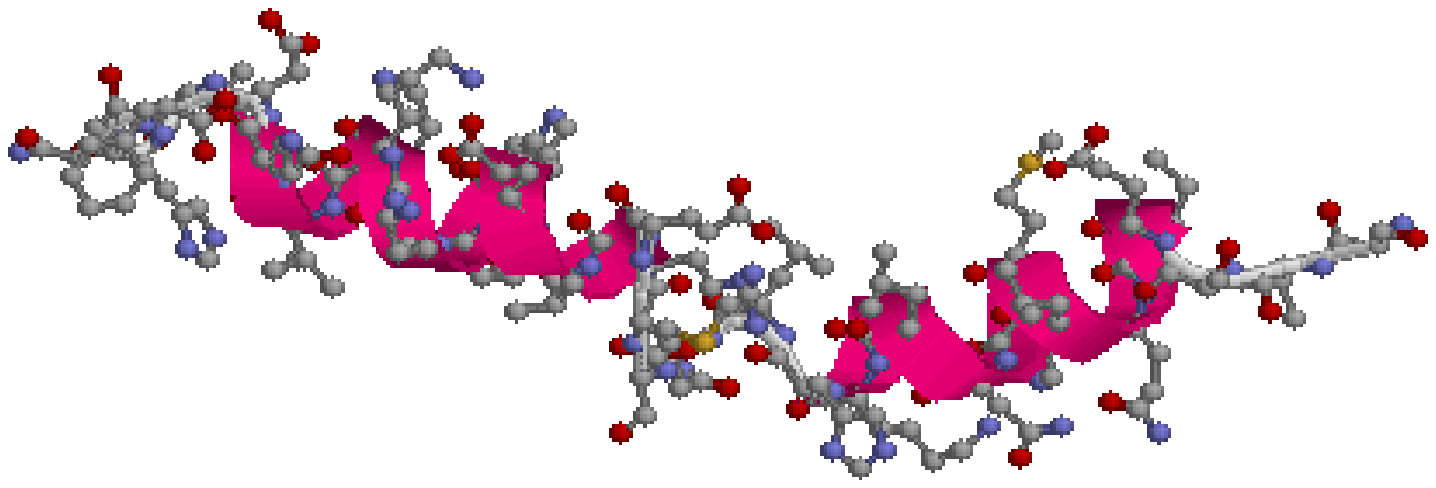}
\caption{}
\end{figure}
\newpage
\thispagestyle{empty}
\cleardoublepage
\begin{figure}
\includegraphics[angle=0,width=0.95\textwidth]{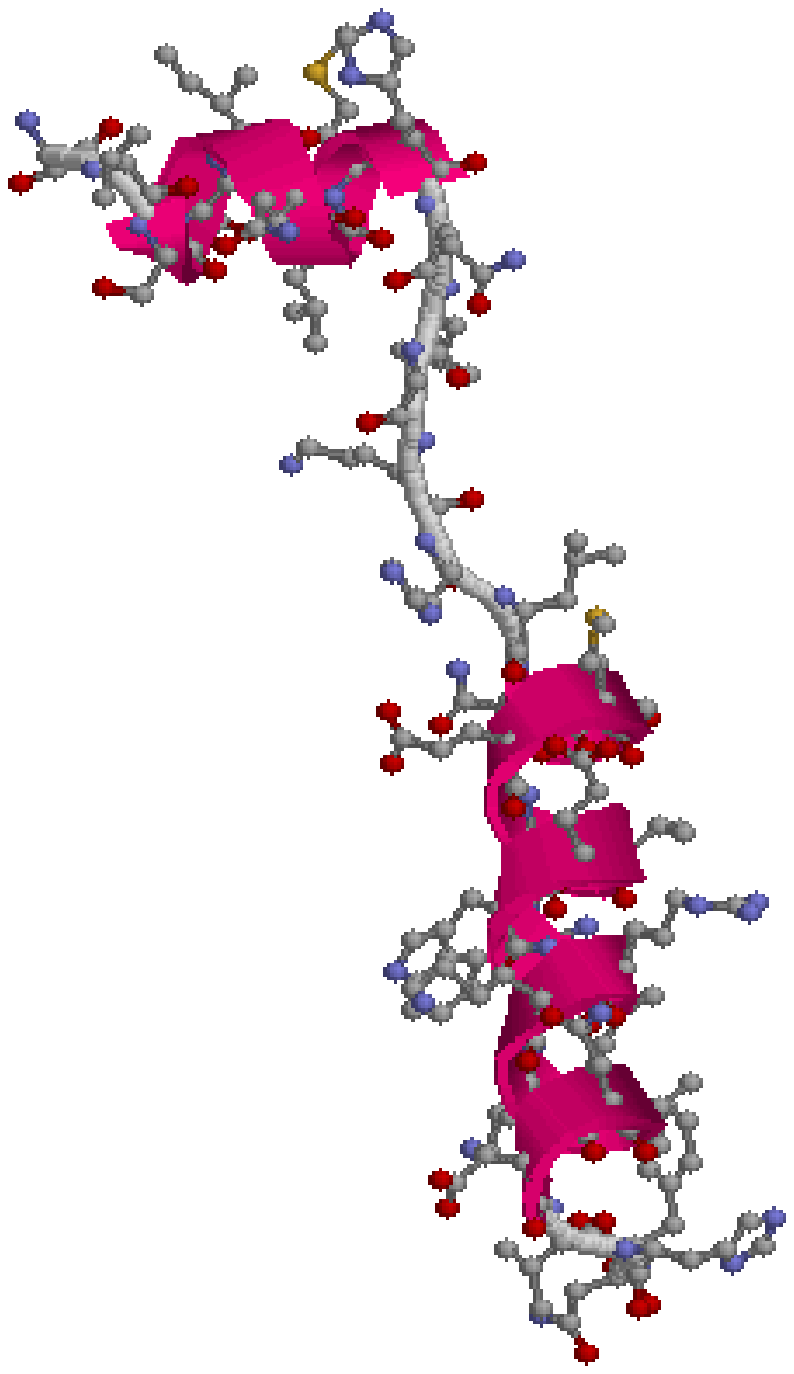}
\caption{}
\end{figure}

\newpage
\thispagestyle{empty}
\cleardoublepage
\begin{figure}
\includegraphics[angle=0,width=0.95\textwidth]{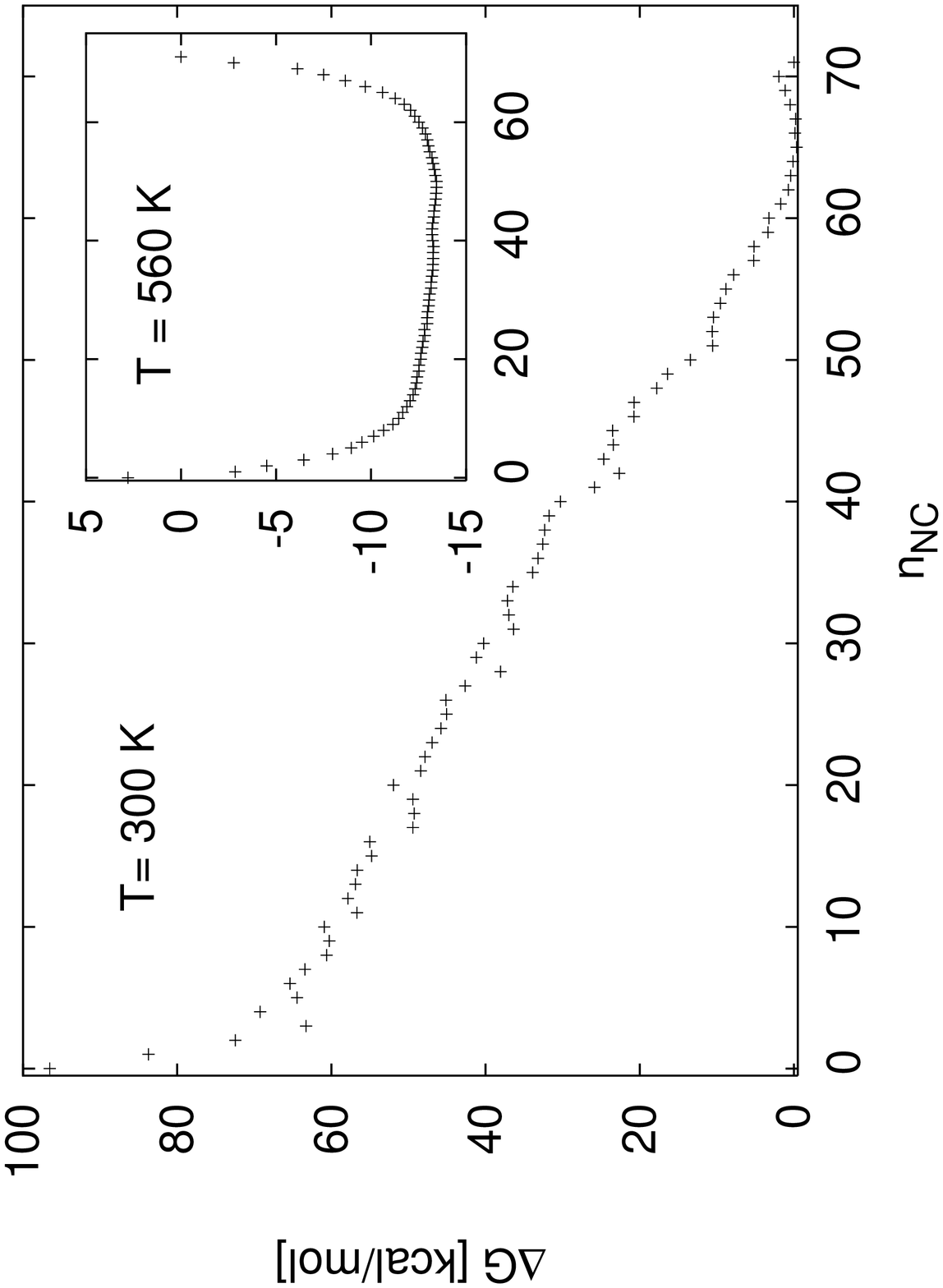}
\caption{}
\end{figure}

\newpage
\thispagestyle{empty}
\cleardoublepage
\begin{figure}
\includegraphics[angle=0,width=0.95\textwidth]{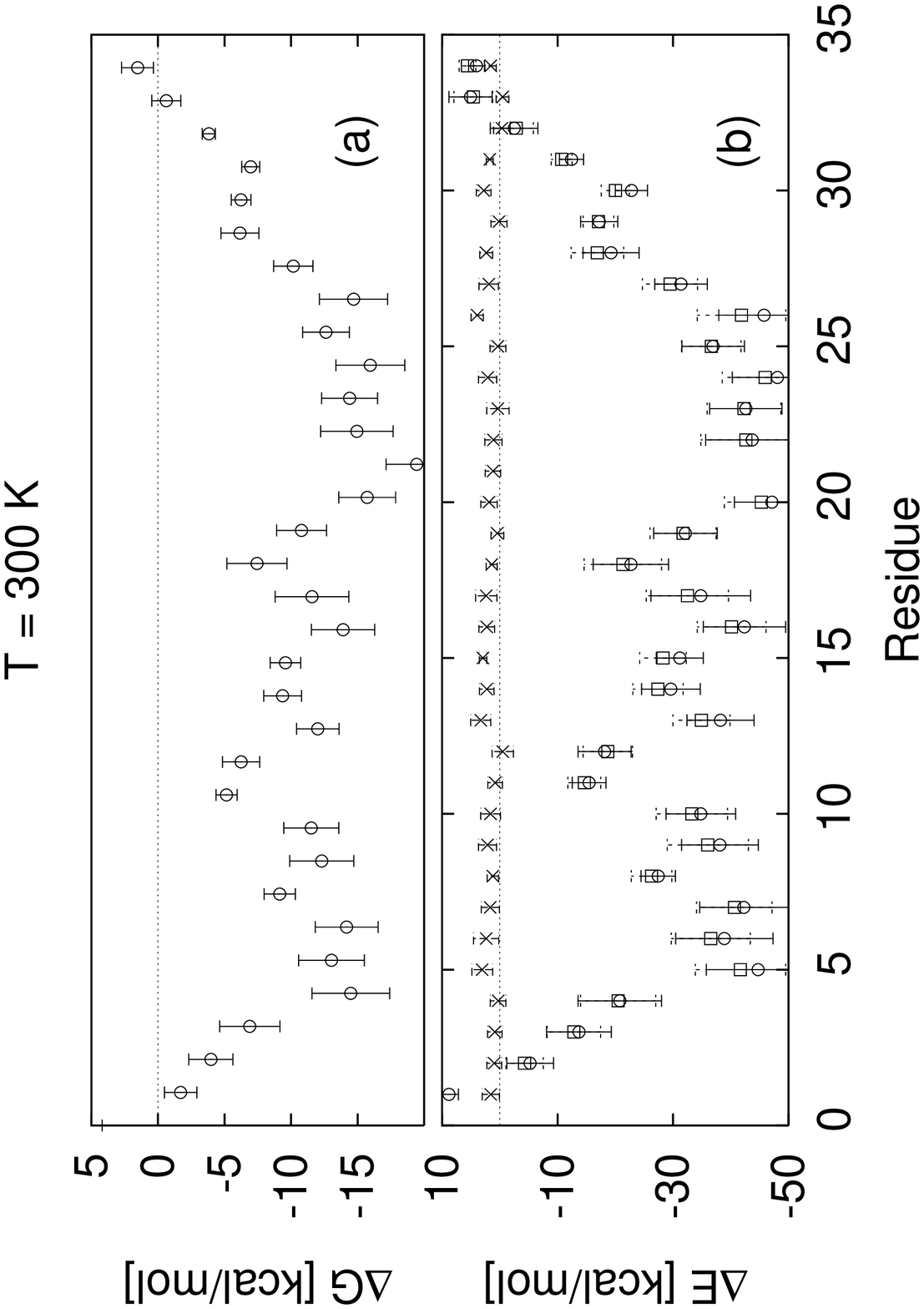}
\caption{}
\end{figure}

\newpage
\thispagestyle{empty}
\begin{figure}
\cleardoublepage
\includegraphics[angle=0,width=0.95\textwidth]{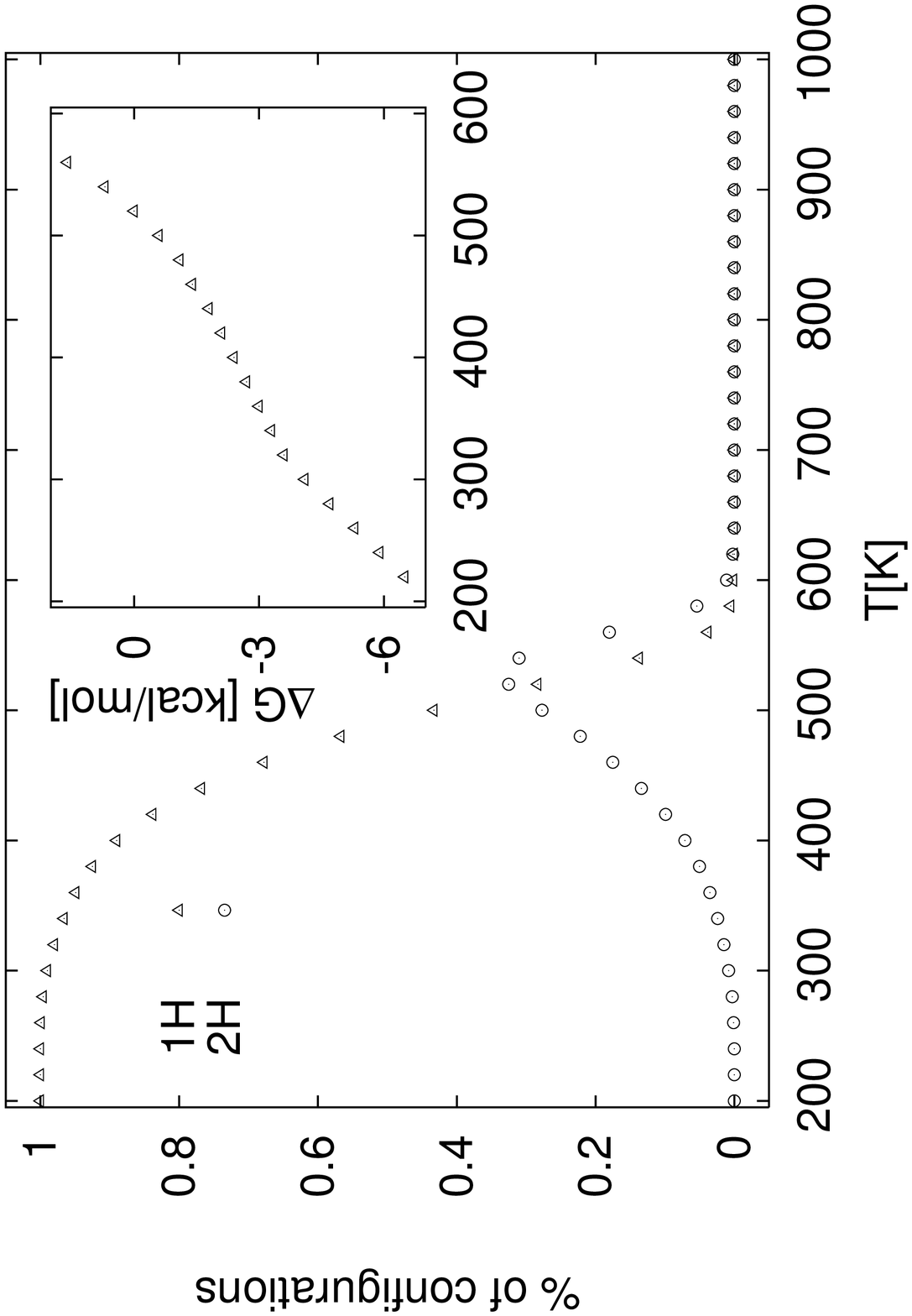}
\caption{}
\end{figure}


\begin{thebibliography}{99}
\bibitem{Calfisch1}  P. Ferrara and A. Caflisch,
                     {\it Proc. Natl. Acad. Sci. USA} {\bf 97}, 10780 (2000).
\bibitem{Kol} Y. Duan and P.A. Kollman, {\it Science}  {\bf 282}, 740 (1998).
\bibitem{Chow} Sh.~Chowdhury, W.~Zhang, Ch.~Wu, G.~Xiong and Y.~Duan,
               {\it Biopolymers} {\bf 68}, 63 (2003).
\bibitem{HO98g} U.H.E.~Hansmann and Y.~ Okamoto,
                {\it Curr. Opin. Struc. Biol.} {\bf 9}, 177 (1999).
\bibitem{Brooks2002} C.L.~Brooks,
                     {\it Accounts of Chemical Res.}, {\bf 35}, 447  (2002).
\bibitem{Elber} A.E. Cardenas and R.~Elber,
                {\it Proteins: Stru. Fu. Gen.}, {\bf 51}, 245 (2003).
\bibitem{H97f} U.H.E.~Hansmann, {\it Chem.~Phys.~Lett.} {\bf 281}, 140 (1997).
\bibitem{Angel} A.E. Garcia and K.Y. Sabonmatsu,
                {\it Proteins: Str. Fu. Gen.}, {\bf 42}, 345 (2001).
\bibitem{LHH} C.-Y. Lin, C.-K.~Hu and U.H.E.~Hansmann,
              {\it Proteins: Str. Fu. Gen.}, {\bf 52}, 436 (2003).
\bibitem{Calfisch2}F. Rao and A. Caflisch,
                   {\it J. Chem. Phys.}, in press.
\bibitem{MyReview} U.H.E.~Hansmann and Y.~ Okamoto, in:
         D. Stauffer (ed),``Annual Reviews in Computational Physics VI'' 
         Singapor: World Scientific (1998), p. 129.
\bibitem{MU}   B.A. Berg  and T. Neuhaus,
             {Phys. Lett.} B {\bf 267}, 249 (1991).
\bibitem{HO1}  U.H.E.~Hansmann and Y.~ Okamoto, {\it J.~Comp.~Chem.}
             {\bf 14}, 1333 (1993).
\bibitem{HO98c}  U.H.E.~Hansmann and Y.~ Okamoto,
                {\it J. Phys. Chem} {\bf  102}, 653 (1998).
\bibitem{HA02}  N.A.~Alves and  U.H.E.~Hansmann,
                {\it J. Chem. Phys.}  {\bf 117},  2337 (2002).
\bibitem{Klaus} W. Klaus, T. Dieckmann, V. Wray, D. Schomburg, 
                E. Wingender and H. Mayer,
                {\it Biochemistry} {\bf  30}, 6936 (1991).
\bibitem{Crystal} L.~Jin, S.L.~Briggs, S.~Chandrasekhar, N.Y.~Chirgadze,
                  D.K.~Clawson, R.W.~Schevitz, D.L.~Smiley, A.H.~Tashjian,
                  F.~Zhang, {\it J. Biol. Chem.} {\bf 275}, 27238 (2000).
\bibitem{MABFR} U.C.~Marx, K.~Adermann, P.~Bayer, W.-G.~Forssmann and P. R{\"o}sch,
                {\it Biochem. Biophys. Res. Comm.} {\bf 267}, 213 (2000).
\bibitem{Potts} J.T.~Potts  Jr, H.M.~Kronenberg, M.~Rosenblatt,
                {\it Adv. Prot. Chem.} {\bf 35}, 323 (1982).
\bibitem{Brommage99} R.~Brommage, C.E.~Hotchkiss, C.J.~Lees, M.W.~Stancill,
                   J.M.~Hock and C.P.~Jerome,
                   {\it J. Clin. Endocrinol. Metab.} {\bf 84}, 3757 (1999).
\bibitem{Okamoto} Y.~Okamoto, T.~Kikuchi, T.~Nakazawa  and H.~Kawai,
                {\it Int. J. Pep. Prot. Res.} {\bf 42}, 300 (1993).
\bibitem{EC3} G.~N\'emethy, K.D.~Gibson, K.A.~Palmer, C.-N.~Yoon, G.~Paterlini, 
              A.~Zagari, S.~Rumsey, H.A.~Scheraga,  
              {\it J.~Phys.~Chem.} {\bf 96},  6472 (1992).
\bibitem{OONS} T. Ooi, M. Obatake, G. Nemethy, H.A. Scheraga,
               {Proc. Natl. Acad. Sci.  USA} {\bf 8}, 3086 (1987).
\bibitem{SMMP}  F.~Eisenmenger, U.H.E.~Hansmann, Sh.~Hayryan, C.-K.~Hu,
               {Comp.~Phys.~Comm.} {\bf 138}, 192 (2001).
\bibitem{FS}  A.M. Ferrenberg and R.H. Swendsen, {Phys.\ Rev.\ Lett.}
             {\bf  61}, 2635 (1988); {Phys. Rev. Lett.} {\bf 63},
             1658(E) (1989), and
              references given in the erratum.
\bibitem{H97} U.H.E.~Hansmann,
              {\it Phys. Rev. E.} {\bf 56}, 6200 (1997).
\bibitem{ELASS} F.~Eisenhaber, P.~Lijnzaad, P.~Argos, C.~Sander and
                M.~Scharf, {\it J. Comp. Chem.} {\bf 16}, 273 (1995).
\bibitem{FANTOM} T.~Schaumann, W.~Braun and K.~Wuthrich,
                 {\it Biopolymers}, {\bf  29},  679 (1990).
\bibitem{HW} U.H.E.~Hansmann and L.~Wille,
              {\it Phys. Rev. Lett.} {\bf 88},  068105 (2002).
\end{thebibliography}
\end{document}